\newcommand{\be}{\begin{equation}}\newcommand{\ee}{\end{equation}}
\newcommand{\bea}{\begin{eqnarray}}\newcommand{\eea}{\end{eqnarray}}
\newcommand{\nn}{\nonumber}\newcommand{\p}[1]{(\ref{#1})}
\newcommand{\lb}[1]{\label{#1}}
\newcommand\s{\scriptscriptstyle}

\newcommand\q{\quad}
\newcommand\qq{\qquad}
\newcommand\ov{\overline}

\newcommand{\vp}{\varphi}
\newcommand{\bvp}{\bar\varphi}
\newcommand{\bnu}{\bar\nu}

\newcommand\tia{\theta^\alpha_i}

\newcommand\btka{\bar{\theta}^{\da k}}

\newcommand{\eka}{\epsilon^\alpha_k}

\newcommand{\beka}{\bar{\epsilon}^{\da k}}

\newcommand{\da}{{\dot{\alpha}}}
\newcommand{\db}{{\dot{\beta}}}

\newcommand{\toa}{\theta_1^\alpha}
\newcommand{\tob}{\theta_1^\beta}
\newcommand{\tta}{\theta_2^\alpha}
\newcommand{\ttb}{\theta_2^\beta}
\newcommand{\tha}{\theta_3^\alpha}

\newcommand{\btoa}{\bar{\theta}^{1\da}}
\newcommand{\btob}{\bar{\theta}^{1\db}}

\newcommand{\btta}{\bar{\theta}^{2\da}}
\newcommand{\bttb}{\bar{\theta}^{2\db}}

\newcommand{\btha}{\bar{\theta}^{3\da}}
\newcommand{\bthb}{\bar{\theta}^{3\db}}

\newcommand\ab{{\alpha\beta}}

\newcommand\adb{{\alpha\db}}
\newcommand\ada{{\alpha\da}}
\newcommand\bdb{{\beta\db}}
\newcommand\bda{{\beta\da}}
\newcommand\padb{\partial_\adb}

\newcommand\pbda{\partial_\bda}

\newcommand\A{{\s A}}

\newcommand{\Dot}{D^1_2}
\newcommand{\Dto}{D^2_1}
\newcommand{\Doh}{D^1_3}

\newcommand{\Dht}{D^3_2}
\newcommand{\Dth}{D^2_3}

\newcommand{\pot}{\partial^1_2}
\newcommand{\pto}{\partial^2_1}
\newcommand{\poh}{\partial^1_3}
\newcommand{\pho}{\partial^3_1}
\newcommand{\pht}{\partial^3_2}
\newcommand{\pth}{\partial^2_3}

\newcommand{\Vot}{V^1_2}
\newcommand{\Vto}{V^2_1}
\newcommand{\Voh}{V^1_3}

\newcommand{\Vht}{V^3_2}
\newcommand{\Vth}{V^2_3}

\newcommand{\Doa}{D^1_\alpha}

\newcommand{\Dta}{D^2_\alpha}

\newcommand{\poa}{\partial^1_\alpha}

\newcommand{\pta}{\partial^2_\alpha}

\newcommand{\pha}{\partial^3_\alpha}

\newcommand{\bDoa}{\ov{D}_{1\da}}

\newcommand{\bDta}{\ov{D}_{2\da}}

\newcommand{\bDha}{\ov{D}_{3\da}}

\newcommand{\bpoa}{\bar{\partial}_{1\da}}

\newcommand{\bpta}{\bar{\partial}_{2\da}}

\newcommand{\bpha}{\bar{\partial}_{3\da}}

\documentclass[12pt]{article}
\def\theequation{\arabic{section}.\arabic{equation}}
\begin{document}
\begin{titlepage}
\begin{flushright}
hep-th/0110074 \\
October, 2001
\end{flushright}
\begin{center}
{\bf  N=3 SUPERSYMMETRIC BORN-INFELD THEORY}\\
\vspace{0.5cm} {\bf E.A. Ivanov${}^{\;1)}$, B.M. Zupnik${}^{\;2)}$}
\vspace{1cm} \\
{\it Bogoliubov Laboratory of Theoretical Physics, Joint Institute for
Nuclear Research, 141980, Dubna, Moscow Region, Russia}\\
\vspace{0.5cm}
\end{center}

\begin{abstract}
We construct an off-shell $N=3$ supersymmetric extension of the
abelian $D=4$ Born-Infeld action starting from the action of
supersymmetric Maxwell theory in $N=3$ harmonic superspace. A crucial new
feature of the $N=3$ super BI action
is that its interaction part contains only terms of
the order $4k$ in the $N=3$ superfield strengths.
The correct component bosonic BI action arises as the result of
elimination of auxiliary tensor field which is present in the
off-shell $N=3$ vector multiplet in parallel with the gauge
field strength. In this new Legendre-type representation, the
bosonic BI action is fully specified by a real function of the
single variable quartic in the auxiliary tensor field. The generic
choice of this function amounts to a wide set of self-dual
nonlinear extensions of the Maxwell action. All of them admit an
off-shell $N=3$ supersymmetrization.
\end{abstract}
\vspace{0.6cm}

\vspace{0.2cm}


duality

\vfill
\noindent{\it E-Mail: \\
1) eivanov@thsun1.jinr.ru \\
2) zupnik@thsun1.jinr.ru}

\end{titlepage}
\section{Introduction}
Supersymmetric extensions of the Born-Infeld (BI) and Dirac-Born-Infeld
actions play the important role in modern string theory, being essential parts
of the worldvolume actions of D$p$-branes \cite{Ts}. A manifestly $N=1$
supersymmetric
$D=4$ BI action was constructed  in Refs. \cite{CF}. Later on, it
was rederived in \cite{BG} within the nonlinear realizations approach as
the Goldstone-Maxwell superfield action describing one of possible patterns
of the partial
spontaneous breaking of $N=2, D=4$ supersymmetry down to $N=1$ supersymmetry.
It was interpreted as a manifestly worldvolume supersymmetric form of
the static-gauge action of the ``space-filling'' super D3-brane.

An interesting and challenging problem is to construct off-shell
super BI actions with manifest (linearly realized) extended
supersymmetries.\footnote{In the approach proceeding from the gauge-fixed
worldvolume D$p$-brane actions \cite{AS} the super BI actions appear in the
component on-shell form, with all involved supersymmetries
realized nonlinearly.} A direct $N=2$ supersymmetrization of the
$D=4$ BI action in terms of the $N=2$ Maxwell superfield strength
was constructed in \cite{Ke}. In \cite{BIK}, a modified action was
proposed, such that it can be interpreted as the
Goldstone-Maxwell superfield action for the partial supersymmetry
breaking $N=4 \rightarrow N=2$ in $D=4$ (and, respectively, as a
static-gauge form of the super D3-brane action in $D=6$, with two
scalar fields of the vector $N=2$ multiplet being the transverse
brane coordinates). No super BI actions with manifest linearly
realized higher $N$ supersymmetries were constructed so far. Only
partial results related to the supersymmetrization of the quartic
term in the $\alpha{}'$ expansion of the BI action (the so called
Euler-Heisenberg (EH) action) were known. An $N=4$ supersymmetric
extension of this term was found in terms of $N=1$ superfields
\cite{Ts} and $N=2$ superfields in the projective \cite{GRo} and
harmonic \cite{Ivunp} $N=2$ superspaces. Accordingly, only
$N=1$ or $N=2$ supersymmetries in these actions are realized
linearly and off-shell. A manifestly $N=3$ supersymmetric
extension of the EH action in the light-cone superspace (lacking
manifest Lorentz invariance) was presented in \cite{BKL}.

In this paper we construct an $N=3$ superextension of the full BI
action in $N=3$ harmonic superspace (HSS). This superspace is a
generalization of $N=2$ HSS \cite{GIK1}, and it was
introduced in \cite{GIK2} to obtain an off-shell unconstrained
superfield formulation of $N=3$ gauge theory (amounting to
$N=4$ gauge theory on shell). The off-shell action of $N=3$ gauge theory has
an unusual form of superfield Chern-Simons-type term and exists entirely due
to a few unique (almost miraculous) peculiarities of $N=3$ HSS. The
opportunity to construct the $N=3$ BI action also amounts to one of such
peculiarities.

The direct $N=1$ and $N=2$ superextensions of the BI action are collections of
separate superfield terms which supersymmetrize each order of the expansion of
the bosonic BI action in powers of the Maxwell field strength. In the $N=3$
harmonic superfield formalism it is easy to construct the 4th order
interaction term from the off-shell $N=3$ superfield strengths defined by
the Grassmann-analytic harmonic gauge potentials. However, the 6th order
superfield term does not exist, though the terms of 8th and higher orders
exist again. Thus, a naive $N=3$ completion of the full variety of the bosonic
nonlinear terms of the BI action seems impossible.

A surprising way around this difficulty is related to the
following unusual feature of the off-shell $N=3$ HSS formalism
\cite{GIK2}. The Grassmann-analytic gauge potentials of $N=3$
gauge theory contain, besides the physical fields including  the
standard gauge potential $A_m$, also an infinite number
of the auxiliary fields. Among them there is an
independent bispinor field $H_\ab$. The correct
bilinear Maxwell term in the component action arises only after
elimination of this field by its algebraic equation of motion.
The $N=3$ superfield strength  contains the combination
$V_\ab={1\over4}[H_\ab+ F_\ab(A)]$ of the auxiliary field
and the gauge field strength.

This off-shell structure of the $N=3$ harmonic superfields
suggests the following way of solving the problem of $N=3$
supersymmetrization of the BI theory. In Ref. \cite{BIK2} it was
noticed that the vector auxiliary components of the off-shell
$N=2$ hypermultiplet in $N=2$ HSS are capable to generate
higher-order terms in the bosonic part of some super $p$-brane action already
from the 4th order superfield term. This observation indicates that in the
presence of tensor auxiliary components the issue of supersymmetric
generalization of given nonlinear bosonic action does not amount to
a straightforward order-by-order reconstruction. We use the auxiliary component
$V_\ab$ as the Legendre-type transform variable for the gauge field strength
$F_\ab(A)$. It turns out that this specific Legendre transform of the standard
bosonic BI action is determined by a real function $E$ of the single variable
$a= V^2\bar{V}^2$ where $V^2=V^\ab V_\ab$. The problem of $N=3$
supersymmetrization is then reduced to the construction of the
superfield terms of the order 4k in the auxiliary field $V_\ab$.
All these terms can be constructed as the appropriate powers of
the off-shell $N=3$ superfield strengths and their spinor
derivatives in the framework of the analytic subspace of $N=3$ HSS.
Thus an off-shell $N=3$ BI action proves to exist despite the
absence of the 6th order self-interaction superfield terms. A
generic function $E(a)$ exhausts the complete set of the $SO(2)$
self-dual nonlinear extensions of the Maxwell action, the BI one
being a special representative of them. All such actions can be
off-shell $N=3$ supersymmetrized.

\setcounter{equation}0
\section{\lb{B0} N=3 harmonic superspace}
\subsection{\lb{B}Constraints of N=3 gauge theory in ordinary \break N=3
superspace}

We start by recapitulating basic facts about the formulation of
$N=3$ supersymmetric gauge theory in the standard $N=3, D=4$
superspace $R(4|12) = \{z^M \}$,
\begin{equation}
z^M=(x^\adb ,~\tia,~\btka)~. \lb{B1}
\ee
Here  $i, k\ldots=1,2,3$ are indices of the  fundamental representations
of the group $SU(3)$, the $R$-symmetry group of the $N=3, D=4$
Poincar\'e superalgebra.

The algebra of spinor derivatives in $R(4|12)$ has the form
\bea
\{ D^k_\alpha, D^l_\beta\}=0~, \quad \{ D^k_\alpha, \bar{D}_{l\db}\}=
-4i\delta^k_l\padb  \quad \mbox{and c.c.}~.\lb{B4}
\eea

The superfield constraints defining the $N=3$ supersymmetric Maxwell theory
are the following gauge-covariantized version of these relations:
\bea
\{\nabla^i_\alpha,\nabla^k_\alpha \}=\varepsilon_\ab
\bar{W}^{ik}~, \quad \{\nabla^k_\alpha,\bar{\nabla}_{l\db}
\} = -4i\delta^k_l\nabla_\adb
\quad \mbox{and c.c.}~, \lb{con3}
\eea
where
\be
\nabla^i_\alpha= D^i_\alpha +A^i_\alpha(z)~,\qq
\bar{\nabla}_{i\da}= \bar{D}_{i\da}
+\bar{A}_{i\da}(z)\lb{B4b}
\ee
and $A^i_\alpha(z), \bar{A}_{i\da}(z)$ are the corresponding spinor gauge
connections.

The Bianchi identities following from \p{con3} produce
the constraints on the covariant superfield strengths
\bea
&& D^i_\alpha \bar{W}^{kl} +D^k_\alpha \bar{W}^{il}=0~,\lb{B5}\\
&&\bar{D}_{i\da}\bar{W}^{kl}={1\over2}(\delta^k_i
\bar{D}_{j\da}\bar{W}^{jl}-
\delta^l_i\bar{D}_{j\da}\bar{W}^{jk}) \quad \mbox{and c.c.}~.\lb{B6}
\eea
They can be shown to reduce the component fields content of the superfield
strengths to the on-shell $N=3$ vector multiplet \cite{So}.

\subsection{\lb{C} Off-shell gauge superfields in N=3 harmonic superspace}

The $SU(3)/U(1)\times U(1)$ harmonic superspace has been
introduced in \cite{GIK2} to construct an off-shell unconstrained
superfield formulation of $N=3$ gauge theories. We quote our conventions
for coordinates and derivatives in $N=3$ HSS in the Appendix A (they
are in essence the same as in ref. \cite{AFSZ,NZ}). Here we
recall the basic elements of abelian $N=3$ gauge theory in $N=3$
HSS using this notation.

The fundamental objects of the abelian $N=3$ gauge theory are
three harmonic gauge potentials living as unconstrained
superfields on the $(4{+}6|8)$-dimensional analytic subspace
$H(4{+}6|8) =\{\zeta, u\}$ of $N=3$ HSS
\bea
&&\Vot(\zeta,u)\equiv
V^{(2,-1)}(\zeta, u) ~,\q\Voh(\zeta,u) \equiv V^{(1,1)}(\zeta,
u)~,\q \Vth\equiv V^{(-1,2)}(\zeta, u)
~,\nn\\
&&\Vot=-\widetilde{(\Vth)}~,\qq\Voh=\widetilde{(\Voh)}~.\lb{C2}
\eea
The definition of the generalized conjugation $\sim $ preserving $N=3$
Grassmann harmonic analyticity and the precise content of the analytic
coordinate  set $\{\zeta, u \}$ are given in Appendix A. We employed here
the double notation in order to emphasize a contact with the original paper
\cite{GIK2}.

The potentials undergo abelian  gauge transformations with a real analytic
parameter $\lambda(\zeta,u)$:  \be
\delta\Vot=i\Dot\lambda~,
\q\delta\Voh=i\Doh\lambda~,\q\delta\Vth=i\Dth\lambda~. \lb{C3}
\ee
The potential $V^1_3$ can be consistently expressed in terms of
the two remaining ones by imposing the conventional constraint
\be
\hat{V}^1_3\equiv \Dot\Vth-\Dth\Vot~.
\ee
Below we shall use the
off-shell formalism with such composite potential $\hat{V}^1_3$.

The free $N=3$ gauge theory action has the following form:
\be
S_2(\Vot,\Vth)=-{1\over4f^2}\int d\zeta(^{33}_{11}) du
\left[\Vth\Doh\Vot+ {1\over2}(\Dot\Vth-\Dth\Vot)^2\right]~,
\lb{freehs}
\ee
where the analytic superspace integration
measure  $d\zeta(^{33}_{11})= d^4x d^8\theta(^{33}_{11})$ is
defined in \p{gme2} and we have introduced the coupling constant
$f$ of dimension $-2$, so that $[V^1_2]= -2$ and the gauge field
strength is dimensionless.
This convention will turn out useful when constructing nonlinear
extensions of \p{freehs}.

Besides an infinite number of gauge components accounted for by the
gauge freedom \p{C3}, the gauge potentials possess an infinite number of
the auxiliary field components. The latter disappear only on the mass shell
defined by the free equations of motion following from \p{freehs}:
\bea
F^{11}_{23}=\Doh \Vot -\Dot \hat{V}^1_3 =0~, \quad
F^{12}_{33}=\Dth \hat{V}^1_3 -\Doh \Vth =0~.\lb{C5}
\eea
These equations, being a sort of harmonic zero-curvature conditions,
imply the on-shell representation of the abelian harmonic
connections through the real non-analytic bridge superfield $v$
\be V^I_K(v)=iD^I_K v~.\lb{bridge} \ee This representation,
together with the $N=3$ Grassmann analyticity conditions for the potentials,
can be used to demonstrate \cite{GIK2} that \p{C5} are equivalent to
the original $R(4|12)$ constraints \p{B5}, \p{B6}.

For our further purposes it will be important to know the full
structure of the bosonic $SU(3)$ singlet sector in the component expansion of
the off-shell analytic potentials $\Vot$ and $\Vth$ in the WZ gauge. A simple
analysis yields \bea
&& v^1_2=\tta\btob A_\adb +i(\theta_2)^2\bar\theta^{1(\da}
\bar\theta^{2\db)} \bar{H}_{\da\db} + i(\theta_2)^2(\bar\theta^1 \bar\theta^2)
C~, \lb{12} \\
&& v^2_3=\tha\bttb A_\adb
-i\theta_2^{(\alpha}\theta_3^{\beta)}(\bar\theta^2)^2H_{\alpha\beta}
-i(\theta_2\theta_3)(\bar\theta^2)^2C, \lb{23} \\
&&\hat{v}^1_3=2\tha\btob A_\adb
-2i(\bar\theta^1 \bar\theta^2)\theta_2^{(\alpha}\theta_3^{\beta)}(H-F)_\ab
+2i(\theta_2\theta_3)\bar\theta^{1(\da} \bar\theta^{2\db)}(\bar{H} -
\bar{F})_{\da\db}\nn\\
&&+4i(\theta_2\theta_3)(\bar\theta^1 \bar\theta^2)C
+ (\theta_2)^2
(\bar\theta^2)^2\tha\btoa\left(\partial^\beta_\da
 H_\ab + \partial^\db_\alpha\bar{H}_{\da\db}\right)~, \lb{singl}
\eea
where $H_{\alpha\beta} = H_{\beta\alpha}~, \bar C = C$ and the spinor
representation for the gauge field strength was used
\bea
&&F_\ab(A)\equiv \partial_{(\alpha}^\db A_{\beta)\db}~,\q
\bar{F}_{\da\db}(A)\equiv \partial^\beta_{(\da}A_{\beta\db)}~,
\lb{defF}\\
&& \partial_\alpha^\db \bar F_{\db\da} -\partial^\beta_\da F_{\beta\alpha}
= 0~. \lb{Bian}
\eea
We observe that the auxiliary dimensionless symmetric tensor and scalar fields
$H_\ab$ and $C$ are present in the off-shell $SU(3)$  singlet
sector in parallel with the gauge potential $A_{\alpha\dot\beta}$
and its covariant field strength. The presence of these auxiliary fields is
the pivotal difference of the off-shell vector $N=3$ multiplet from the
$N=2$ one arising from the WZ gauge of the analytic harmonic $N=2$ gauge
connection \cite{GIK1}. As we shall see soon, the fields $H_{\alpha\beta},
 \bar H_{\da\db}$ play a crucial role in constructing $N=3$ supersymmetric BI
action.

It should be emphasized that the above $SU(3)$ singlet fields
are  components of the infinite-dimensional off-shell $N=3$ gauge
supermultiplet. But all other bosonic fields in the WZ gauge have
a non-trivial $SU(3)$ assignment. In what follows we shall be interested in
the pure Maxwell part of the component form of the action \p{freehs}, so for
us it will be enough to know just the $SU(3)$ singlet sector \p{12} -
\p{singl} of the analytic gauge potentials.

It is easy to explicitly evaluate contributions of two terms in
the action \p{freehs} to the $SU(3)$ singlet part of the
component form of this action (up to surface terms)
\bea
-{1\over4f^2}\int d\zeta(^{33}_{11}) du\, v^2_3\Doh v^1_2
&\Rightarrow &
-{1\over8f^2}\int d^4x\, (H F + \bar H\bar F)~,\\
-{1\over8f^2}\int d\zeta(^{33}_{11}) du \,(\hat{v}^1_3)^2 &\Rightarrow &
{1\over16f^2}\int d^4x\, [\,H^2+\bar{H}^2- 4\,(HF +\bar{H}\bar{F})\nn\\
&& +\,F^2+ \bar{F}^2+ 8 C^2\,]~,
\eea
where the definition \p{gme2} and the following notation
\bea
&&F^2=F^\ab F_\ab~,\q H^2=H^\ab H_\ab~,\q FH=F^\ab H_\ab~,\\
&&\bar{F}^2=\bar{F}^{\da\db} \bar{F}_{\da\db}~,\q
\bar{H}^2=\bar{H}^{\da\db} \bar{H}_{\da\db}~,\q
\bar{F}\bar{H}=\bar{F}^{\da\db} \bar{H}_{\da\db}
\eea
were used.

Thus the gauge field part of the off-shell super $N=3$
Maxwell component Lagrangian is
\be
L_2(F,H,C)={1\over16f^2}[\,H^2+\bar{H}^2 -6\,(\bar{H}\bar{F}+ H
F)+ F^2+\bar{F}^2 + 8 C^2 \,]~. \lb{LFH}
\ee
Eliminating
the auxiliary fields $H_{\alpha\beta}, \bar H_{\da\db}, C$ by
their algebraic equations of motion
\be
H_{\alpha\beta} =3\,F_{\alpha\beta}~, \q \bar H_{\da\db} = 3\,
\bar F_{\da\db}~,\q C = 0~,
\ee
we arrive at the
standard Maxwell action
\be
L_2(F) = - {1\over2f^2}(F^2 + \bar F^2) = - {1\over 4f^2}{\cal F}^{mn}
{\cal F}_{mn}~, \lb{Maxstand}
\ee
where ${\cal F}_{mn}=\partial_m
A_n-\partial_n A_m$ and the precise relation of the spinor and vector
notations is given in Appendix B. We
shall see soon that the appropriate starting point for the construction of
$N=3$ BI action is just the $N=3$ supersymmetry-inspired form \p{LFH}
of the Maxwell action, with the properly redefined tensor auxiliary fields.

To this end, let us construct the analytic superfield strengths
for the $N=3$ gauge theory. Like in the $N=2$ gauge theory in
$N=2$ HSS \cite{Zu2}, one firstly defines the non-analytic abelian
connections via the harmonic zero-curvature equations
\bea
\Dot\Vto -\Dto \Vot =0~,\q \Dth \Vht -\Dht \Vth=0~,\lb{C9}
\eea
where
$V^3_2 = -\widetilde{V^2_1}, \,\delta V^3_2 = iD^3_2 \lambda,\,
\delta V^2_1 = iD^2_1 \lambda$ and the explicit form of the
harmonic derivatives is given in Appendix A. Then
the mutually
conjugated Grassmann-analytic off-shell superfield strengths of the $N=3$
Maxwell theory are constructed as follows \cite{NZ}:
\be
W_{23}={1\over4}(\bar{D}_3)^2V^3_2~, \q
\bar{W}^{12}=-{1\over4}(D^1)^2V^2_1~,
\lb{2312}
\ee
i.e., quite analogously to the construction of superfield strengths in $N=2$
HSS.

These off-shell superfield strengths satisfy the following Grassmann
analyticity conditions:
\bea
\bDta W_{23}=\bDha W_{23}=\Doa W_{23}=0~, \q
\Doa\bar{W}^{12}=\Dta\bar{W}^{12}=\bDha\bar{W}^{12}=0 \lb{C31b}
\eea
and harmonic differential conditions
\be
\Dth W_{23}=0\q,~\Dot\bar{W}^{12}=0~.\lb{kin}
\ee
They can be checked using the analyticity of the
basic gauge potentials, the relations \p{C9}, and
the properties like
$$
D^2_1 f_2 = 0 \;\Rightarrow \; f_2 = 0~,
$$
which is valid for any $SU(3)$ harmonic function with at least
one lower-case index 2 (such relations can be easily proved
in the central basis, where harmonic derivatives are short).
It is also straightforward to check gauge invariance
of the superfield strengths \p{2312}.

Free  equations of motion for the harmonic potentials (\ref{C5}) yield
the  on-shell harmonic-analyticity equations for the superfield strengths
\cite{AFSZ}
\be
V^I_K=iD^I_K v~ \Rightarrow~\Dot
W_{23}=0\q,~\Dth\bar{W}^{12}=0~.\lb{free2}
\ee
Together with
\p{kin}, these equations imply that on shell and in the central
basis
\be W_{23}=u_2^ku_3^lW_{kl}(z^M)~,\q
\bar{W}^{12}=u^1_ku^2_l\bar{W}^{kl}(z^M)~, \lb{cbrep1}
\ee
where
the superfield strengths $W_{kl}, \bar{W}^{kl}$ satisfy the
original constraints \p{B5}, \p{B6}. An important consequence of the dynamical
harmonic constraints \p{free2} are the following on-shell conditions:
\be
(D^2)^2W_{23}=(D^3)^2W_{23} =(\bar{D}_1)^2 W_{23}= (D^3)^2\bar{W}^{12}
=(\bar{D}_1)^2\bar{W}^{12}= (\bar{D}_2)^2 \bar{W}^{12}=0~.\lb{short}
\ee

We shall need the full off-shell $SU(3)$ singlet component structure of
$W_{23},\bar{W}^{12}$. In order to find it one should firstly solve the
harmonic equations \p{C9} for the corresponding parts of the
non-analytic harmonic connections ($v^2_1$ and $v^3_2$), assuming the
ansatz \p{12}, \p{23} for the analytic connections $V^1_2$ and
$V^2_3$. Using the property $(D^1_2)^2v^1_2=0 $, one
obtains the following exact representation for $v^2_1$ in terms
of $v^1_2$:
\be
v^2_1={1\over2}(D^2_1)^2 v^1_2
-{1\over12}(D^2_1)^3 D^1_2v^1_2~.\lb{reprv21}
\ee
The expression
for $v^3_2$ can be obtained by the $\sim$ conjugation of $v^2_1$.
Substituting \p{reprv21} into \p{2312}, we get the corresponding
representation for $\bar w^{12}$:
\be
\bar{w}^{12}\equiv-{1\over4}(D^1)^2v^2_1 =-{1\over4}(D^2)^2v^1_2
+{1\over8}(D^2)^2\Dto\Dot v^1_2~.
\ee
Once again, $w_{23}$ can be
recovered by the $\sim $ conjugation.

The relevant explicit expressions are
\bea
v^2_1 &=& - \theta_1^\alpha \bar\theta^{2\db} A_\adb
-i(\bar\theta^2)^2\theta_1^{\alpha}\theta_2^{\beta}\left(F_{\alpha\beta}
+ \epsilon_{\alpha\beta}C\right) +i
(\theta_1)^2\bar\theta^{1\da}\bar\theta^{2\db}
\left( \bar H_{\da\db} +\bar F_{\da\db}\right) \nn \\
&& -(\bar\theta^2)^2(\theta_1)^2\theta_2^{\alpha}\bar\theta^{1\db}
\partial_{\alpha}^{\da}\left( \bar H_{\da\db} +\bar F_{\da\db}\right)~,
\lb{21singl} \\
v^3_2 &=& - \theta_2^\alpha \bar\theta^{3\db} A_\adb
+i(\theta_2)^2\bar\theta^{2\da}\bar\theta^{3\db}\left(\bar F_{\da\db}
- \epsilon_{\da\db}C\right) -i(\bar\theta^3)^2\theta_2^{\alpha}
\theta_3^{\beta}\left( H_{\alpha\beta}+ F_\ab\right) \nn \\
&& + (\bar\theta^3)^2(\theta_2)^2\theta_3^{\alpha}\bar\theta^{2\db}
\partial_{\db}^{\beta}\left( H_{\alpha\beta}+ F_\ab\right)~, \lb{32singl} \\
w_{23} &=& i\theta_2^{\alpha}\theta_3^{\beta}\left( H_\ab+ F_\ab\right)
- (\theta_2)^2\theta_3^{\alpha}\bar\theta^{2\db}
\partial_{\db}^{\beta}\left( H_\ab+ F_\ab\right)
~, \lb{23single} \\
\bar w^{12} &=& i\bar\theta^{1\da}\bar\theta^{2\db}
\left(\bar H_{\da\db} +\bar F_{\da\db}\right)
+(\bar\theta^2)^2\theta_2^{\alpha}\bar\theta^{1\db}
\partial_{\alpha}^{\da}\left(\bar H_{\da\db} +\bar F_{\da\db}\right)
~. \lb{12single}
\eea
One can directly check that $w_{23}, \bar w^{12}$ on their own
obey the off-shell conditions \p{C31b} and \p{kin}.

One observes two distinguished features of $W_{23}$ and $\bar W^{12}$. Firstly,
they do not include the scalar auxiliary field $C(x)$ which is present in the
free $N=3$ gauge theory action \p{freehs}, \p{LFH}. \footnote{This field
enters some other superfield strengths which are of no relevance for our
purpose of constructing a minimal $N=3$ extension of the BI action.} Secondly,
the gauge field strengths appear inside them only in a fixed combination with
the tensor auxiliary fields $H_{\alpha\beta},\bar H_{\da\db}$. Thus, the gauge
field strengths can be fully removed from $W_{23}$ and $\bar W^{12}$ by
redefining the auxiliary fields \bea
&& H_{\alpha\beta} \; \Rightarrow \; V_{\alpha\beta} = {1\over 4}\left(
H_{\alpha\beta} + F_{\alpha\beta} \right)~, \;
\bar H_{\da\db} \; \Rightarrow \; \bar V_{\da\db} = {1\over 4}\,\left(
\bar H_{\da\db} + \bar F_{\da\db} \right)~, \lb{redef} \\
&& W_{23} = 4i\,\theta^{(\alpha}_2\theta^{\beta)}_3V_\ab+\ldots~, \q
\bar{W}^{12} = 4i\,\bar\theta^{1(\da} \bar\theta^{2\db)}\bar{V}_{\da\db}
+\ldots~,\lb{gs3b}\\
&& (W_{23})^2=
 4(\theta_2)^2(\theta_3)^2V^2+\ldots\q(\bar{W}^{12})^2= 4(\bar\theta^1)^2
(\bar\theta^2)^2\bar{V}^2+\ldots~. \lb{gs1}
\eea
The free Maxwell Lagrangian \p{LFH} (with $C=0$), being rewritten
through $V_{\alpha\beta}, \bar V_{\da\db}$, reads
\be
L_2(F,H,0)\equiv
B_2(F,V)={1\over f^2}[\,V^2+ \bar{V}^2- 2\,(V F+\bar{V}\bar{F})
+{1\over2}(F^2+\bar{F}^2)\,]~. \lb{auxfree}
\ee
The algebraic equations of motion for  $V_{\alpha\beta}, \bar V_{\da\db}$
giving rise to the standard Lagrangian \p{Maxstand} are simply
\be
V_{\alpha\beta} =  F_{\alpha\beta}~, \q \bar V_{\da\db} =
 \bar F_{\da\db}~. \lb{Veqs}
\ee

From the above discussion one infers two important properties of
the off-shell description of $N=3$ gauge theory in $N=3$ HSS
having no direct analogs in the $N=1$ and $N=2$ cases. First, the
free Maxwell component Lagrangian appears in the unusual forms
\p{LFH} or \p{auxfree}, while its standard form is recovered only
after eliminating the auxiliary fields $V_{\alpha\beta}, \bar
V_{\da\db}$ by their linear algebraic equations of motion
\p{Veqs}. Secondly, the off-shell superfield strengths contain just
these tensor auxiliary fields, but not the ordinary gauge field
strengths $F_{\alpha\beta}, \bar F_{\da\db}$.

These surprising features suggest a non-standard approach to
constructing nonlinear and non-polynomial superextensions of the
off-shell $N=3$ Maxwell theory. One should start by modifying
\p{LFH} by proper terms which are nonlinear (and/or
non-polynomial) in the auxiliary fields $V_{\alpha\beta}, \bar
V_{\da\db}$, such that nonlinearities in $F_{\alpha\beta}, \bar
F_{\da\db}$ are regained as the result of eliminating these
auxiliary fields by their {\it nonlinear} equations of motion.
Then one can hope to $N=3$ supersymmetrize the terms nonlinear in
$V_{\alpha\beta}, \bar V_{\da\db}$ with the help of the above
superfield strengths $W_{23}, \bar W^{12}$ which contain just
these auxiliary fields. In the next Sections we shall show that
in this way the BI action as well as a wide class of self-dual
extensions of the Maxwell action can be $N=3$ supersymmetrized.

\setcounter{equation}0
\section{New Legendre-type representation of the \break
 Born-Infeld action and self-dualities}

Let us introduce the notation
\be
\varphi = F^2~, \quad \bar\varphi  = \bar
F^2~, \quad X(\vp , \bvp) \equiv
(\vp+\bvp)+(1/4)(\vp-\bvp)^2~. \lb{not}
\ee
In terms of
these variables the standard BI Lagrangian has the following form:
\bea L_{BI}(F,\bar{F}) &=&{1\over f^2}\left[1-
\sqrt{-\mbox{det}(\eta_{mn} + {\cal F}_{mn})}\right] \equiv
{1\over f^2}\left[1-Q(\vp,\bvp)\right]~, \lb{biact}
\\
Q(\vp,\bvp) &=& \sqrt{1+X}=1+{1\over2}X-{1\over8}X^2+{1\over16}X^3
-{5\over128}X^4+\ldots\lb{root}\\
&=& \,1+{1\over2}(\vp+\bvp)-{1\over2}\vp\bvp
+{1\over2^2}\vp\bvp(\vp+\bvp)-{1\over2^3}\vp\bvp(3\vp\bvp+
\vp^2+\bvp^2)+O(\vp^5)\nn
\eea
(one should make use of Eqs.\p{vecsp}
of the Appendix B).

As was already mentioned, our ultimate aim is to find a nonlinear
extension of the free Maxwell Lagrangian in the $N=3$
supersymmetry-inspired form $B_2(F, V)$, eq. \p{auxfree}, such
that this extension becomes the BI Lagrangian \p{biact}
after eliminating the auxiliary fields $V_{\alpha\beta}, \bar
V_{\dot\alpha\dot\beta}$ by their algebraic equations of motion.
By Lorentz covariance, such a nonlinear Lagrangian should have
the following general form: \bea
B(F,V) &=& B_2(F,V) +{1\over f^2} E(V^2,\bar{V}^2) \nn \\
       &=& {1\over f^2}[\nu+\bnu -
2(V F +\bar{V}\bar{F})+{1\over2}(\vp+\bvp)+ E(\nu,\bnu)]~,
\lb{legact}
\eea
where $\nu \equiv V^2, \bnu \equiv \bar V^2$ and
$E(\nu, \bnu)$ is a real function to be determined. The
nonlinear generalization of the free equations \p{Veqs} reads
\be
\partial B(F, V)/\partial V_\ab = 0 \; \Rightarrow \;
V_\ab =  F_\ab\frac{1}{[1+ \partial E(\nu,\bnu)/\partial\nu]}
\equiv  F_\ab N(\nu, \bnu) \lb{FV}
\ee
(with its conjugate).
Further, \p{FV} implies
\bea
&& \nu = \vp\,N^2(\nu,\bnu)~, \quad \bnu = \bvp\, \bar N^2(\nu, \bnu) \;
\Rightarrow \; \nu = \nu(\vp, \bvp)~, \;\; \bnu = \bnu(\vp, \bvp)~, \lb{xz} \\
 && N\left(\nu(\vp, \bvp),\bnu(\vp, \bvp)\right)
 \equiv G(\vp, \bvp)~. \lb{defG}
\eea

The function $ G(\vp, \bvp)$ can be found from the basic
requirement that  \p{legact} coincides with \p{biact}
after elimination of $V_\ab, \bar V_{\dot\alpha\dot\beta}$
\be
\nu+\bnu - 2(VF+\bar{V}\bar{F})+{1\over2}(\vp+\bvp)+ E(\nu,\bnu) =
f^2L_{BI}(\vp, \bvp) = 1 - Q(\vp, \bvp)~. \lb{base1}
\ee
After
substituting the expression \p{FV} and its conjugate for $V_{\alpha\beta},
\bar V_{\dot\alpha\dot\beta}$ and making use of the definition \p{defG}, this
condition can be rewritten as
\be
{1\over2}(\vp+\bvp)- 2(\vp G+\bvp \bar{G}) +\nu+\bnu + E(\nu,\bnu) =
f^2L_{BI}(\vp, \bvp) = 1 - Q(\vp, \bvp)~. \lb{base2}
\ee
Differentiating it with respect to $\vp$ and using
the relations
$$
\frac{\partial \nu}{\partial \vp} = G^2 +2\vp G\frac{\partial G}{\partial\vp}
~, \quad
\frac{\partial\bnu}{\partial\vp} = 2\bvp\bar G\frac{\partial \bar G}
{\partial\vp}~,
$$
which follow from \p{xz}, one obtains the simple expression for
$G(\vp, \bvp)$:
\be
G(\vp, \bvp)= {1\over 2}\left(1-2f^2\frac{\partial
L_{BI} }{\partial\vp}\right)= {1\over 2}\left\{1 + {1\over Q(\vp, \bvp)}
\left [1 + {1\over2}(\vp - \bvp)\right]\right\}~. \lb{exprG}
\ee
A useful corollary of this representation is
\be
G+\bar{G}= 1+ {1\over Q}~.\lb{explic}
\ee
The relation inverse to \p{exprG}
reads
\be
\vp=2\,\bar G\, \frac{1-\bar G }{[1 -(G +\bar G)]^2}~.\lb{zG}
\ee

Our aim is to find $E$ as a function of the variables $\nu = V^2,
\bnu = \bar V^2$. As the first step, one expresses $\nu, \bnu$
in terms of $G$ and $\bar G$, using \p{xz} and \p{zG}
\be
\nu= \vp\, G^2 = 2\,\bar G G^2 \frac{1 -\bar G}{[1 -(G +\bar G)]^2}~.
\lb{xG}
\ee
Introducing
\be
t \equiv \frac{G\bar G}{1 -(G+\bar
G)}~, \lb{deft}
\ee
one finds that $t$, as a consequence of
\p{xG}, satisfies the following quartic equation:
\be
t^4 + t^3-{1\over4} \nu\bnu = 0~. \lb{quart}
\ee
It allows one to
express $t$ in terms of $a \equiv \nu\bnu$
\be
t(a)=-1-\frac{a}{4}+\frac{3a^2}{16}-\frac{15a^3}{64}+
\ldots~.
\ee
Of course, one can write a closed expression for
$t(a)$ as the proper solution of \p{quart}, but we do not
present it here in view of its complexity. The next (and last)
step is to find $E(\nu, \bnu)$.
Taking into account the explicit expressions \p{zG}, \p{xG} and
\p{explic} and substituting all this into \p{base2}, one finally
finds a simple expression for $E(\nu, \bnu)$ through the real
variable $t$ (and much more involved expression in terms of $a
= \nu\bnu$):
\be
E(a)\equiv E[t(a)]=2[2t^2(a)+3t(a)+1] =
\frac{a}{2}-\frac{a^2}{8}+ \frac{3a^3}{32}+\ldots~.
\lb{explE}
\ee
The remarkable property of  $E$ which is most important for further
consideration is that it is a function of the {\it single} real
variable $a = V^2\bar V^2$ which is {\it quartic} in the
auxiliary fields. Thus only terms $\sim V^{2n}\bar V^{2n}$ can appear
in the power expansion of $E(V^2, \bar V^2)$.

Using the representation \p{explE} we can find
$$
\frac{\partial E}{\partial\nu} = 2\, (3 +4t)\,
\frac{\partial t}{\partial\nu}~.
$$
On the other hand, from eq. \p{quart},
$$
\frac{\partial t}{\partial\nu} =
\frac{\bnu}{4 \,(3 + 4t)t^2} = {1\over
2}\frac{1 - G}{(3 + 4t)G}~,
$$
where we have used eqs. \p{xG} and \p{deft}. Thus
\be
\frac{\partial E}{\partial\nu} = G^{-1} -1\lb{EG}
\ee
in agreement with \p{FV} and \p{defG}.

Finally, let us show that the substitution of an {\it arbitrary}
function of the variables $t$ or $a = V^2\bar V^2$ for $E(V^2, \bar V^2)$
into \p{legact} gives rise, upon eliminating $V_\ab, \bar
V_{\dot\alpha\dot\beta}$, to a general set of {\it self-dual}
nonlinear extensions of the Maxwell Lagrangian.

As the first step in proving this, let us note that for an arbitrary
nonlinear extension $L(F,\bar F)$ of the Maxwell action one can
always pick up the appropriate function $E(V^2, \bar V^2)$ such
that
\be
B(F,V(F)) = L(F, \bar F)~. \lb{legact1}
\ee
Further,
after adding a Lagrange multiplier term to such general
$B(F,V)$,
\bea
&& B(F,V) \quad \Rightarrow \quad \tilde B(F,V,P) = B(F,V) +
{i\over f^2}\left(P_{\alpha\beta}F^{\alpha\beta} - \bar
P_{\dot\alpha\dot\beta}\bar
F^{\dot\alpha\dot\beta}\right)~, \lb{modB} \\
&& P_\ab(B)\equiv \partial_{(\alpha}^\db B_{\beta)\db}~,\;
\bar{P}_{\da\db}(B)\equiv \partial^\beta_{(\da}B_{\beta\db)}~,
\lb{dualStr}
\eea
it becomes possible to reproduce the Bianchi
identity \p{Bian} for the field strengths $F_\ab, \bar F_{\da\db}$
and, hence, their standard representation through the Maxwell
potential $A_{\alpha\dot\alpha}$ (Eqs. \p{defF}), by varying \p{modB} with
respect to the unconstrained Lagrange multiplier
$B_\ada$ (the dual gauge potential). On the other hand,
since $F_\ab, \bar F_{\da\db}$ are now off-shell
unconstrained, one can trade them for the dual gauge field strengths
$P_\ab, \bar
P_{\da\db}$ using their algabraic equations of motion:
$$
F_\ab = 2V_\ab - i P_\ab \quad \mbox{and c.c.}~.
$$
It is easy to show that after substituting this expression back
into $\tilde B(F,V P)$ the latter becomes
\be
\tilde B(F, V, P)
\quad \Rightarrow \quad B_2(P, -iV)+ {1\over f^2}E(V^2,
\bar{V}^2) \equiv B_2(P, \tilde V)+ {1\over f^2}E(-\tilde V^2,
-\bar{\tilde{V}}^2)~. \label{mediate}
\ee
The self-duality means
that the final Lagrangian (after elimination of $V_{\alpha\beta}$
and $\bar V_{\dot\alpha\dot\beta}$) has the same form in
terms of $P_{\alpha\beta}$ and $\bar P_{\dot\alpha\dot\beta}$ as
the original one \p{legact1} in terms of $F_{\alpha\beta}$ and
$\bar F_{\dot\alpha\dot\beta}$. From \p{mediate} it is clear that
the necessary and sufficient condition for such  a self-duality is
that the function $E$ is even with respect to its both arguments,
\be E(V^2, \bar V^2) = E(-V^2, -\bar V^2)~.\lb{discrCond} \ee
Obviously, this is valid for an arbitrary function of $a=V^2\bar
V^2$, which proves the self-duality of the corresponding class of
nonlinear actions, including the BI action.

The ``discrete'' self-duality just discussed is sometimes called
``self-duality by Legendre transformation'' \cite{KT}. There
exists another type of self-duality which can be called
$SO(2)$-duality. It holds essentially on shell and can be
formulated as the property of covariance with respect to $SO(2)$
transformations mixing up the Bianchi identities for
$F_{\alpha\beta}$, $\bar F_{\dot\alpha\dot\beta}$ with the equations
of motion associated with the Lagrangian $L(F, \bar F)$. The
differential condition which singles out the Lagrangians
$L(F,\bar F)$ revealing such a type of self-duality is as follows
(for details, see \cite{GZ,KT}):
\be
F^2 - \bar F^2 + P^2 -\bar P^2
= 0~, \lb{condloc1}
\ee
where now
\be P_{\alpha\beta} \equiv
if^2\frac{\partial L(F,\bar F)}{\partial F^{\alpha\beta}}~,
\quad P^2 \equiv P^{\alpha\beta}P_{\alpha\beta}~, \;\;\bar P^2
\equiv \bar P^{\dot\alpha\dot\beta}P_{\dot\alpha\dot\beta}~.
\ee

In order to find the restrictions which this kind of self-duality imposes
on the function $E(V^2, \bar{V}^2)$, let us again start from the
representation \p{legact1}, with $L(F,\bar F)$ being general and unspecified
for the moment. Differentiating this identity with respect to
$F^{\alpha\beta}$ and taking account of the relations \p{base1},
\p{base2} with $L(F,\bar F)$ instead of $L_{BI}$, as well as of
the general relation \p{EG}, one obtains  \be
P_{\alpha\beta}(F) = -2i V_{\alpha\beta}(F) +
iF_{\alpha\beta}~.
\ee
Substituting this back into \p{condloc1}
brings the latter into the form
\be
V^2 - \bar V^2 - (V F - \bar
V\bar F) = 0~.
\ee
Using the same general relations once again,
after some algebra one finds that the latter condition is reduced to
the following {\it linear} differential constraint on the function
$E(\nu, \bnu)$:
\be
\nu\frac{\partial E(\nu, \bnu)}{\partial\nu} =
\bnu\frac{\partial E(\nu, \bnu)}{\partial \bnu}~.
\ee
After passing
to the variables $a= \nu\bnu$, $b = \nu + \bnu$, this condition becomes
\be
\frac{\partial E(a, b)}{\partial b}= 0 \quad \Rightarrow \quad E = E(a)=
E(\nu\bnu)~. \label{condfin}
\ee
Thus we come to the surprising result that the {\it
whole} class of nonlinear extensions of the Maxwell action admitting the
$SO(2)$ self-duality is parametrized by an arbitrary  real
function of one argument $E(\nu\bnu)$ (the only natural
restriction is $E(0) = 0$, which implies the standard Maxwell
action in the limit of vanishing self-interaction).

A similar conclusion has been made in \cite{GR} in a different
context. The authors of \cite{GR} have reduced the $SO(2)$ self-duality
equation \p{condloc1} to a nonlinear differential equation for
$L(F, \bar F)$ and have found that its perturbative solution is
specified by some arbitrary function of the single variable
$\sqrt{F^2\bar{F}^2}$ (in our notation). Our consideration clearly
demonstrates that the class of actions which reveal the
``discrete'' self-duality is wider than that of the $SO(2)$
self-dual ones: the functions \p{condfin} form a subclass of
\p{discrCond}. The BI action obviously respects both types
of self-duality.

\setcounter{equation}0
\section{\lb{D} N=3 Born-Infeld action and its generalizations}

The problem of constructing a manifestly $N=3$
supersymmetric superfield action which would yield, in the bosonic sector, the
previously elaborated $F,V$ form of the BI action amounts to setting up a
collection  of superfield monomials which extend the appropriate terms in
the power expansion of the function $E(V^2\bar V^2)$ defined in \p{explE}. This
procedure is in a sense analogous to the construction  of the $N=1$ and $N=2$
superfield BI actions \cite{CF,Ke} (see also \cite{Ts}). An essential
difference is, however, that in our case we are led to supersymmetrize the
powers of the auxiliary fields $V_{\alpha\beta}, \bar
V_{\dot\alpha\dot\beta}$, while in the $N=1$ and $N=2$ cases the powers of
$F_{\alpha\beta}, \bar F_{\dot\alpha\dot\beta}$ in the expansion of the
standard form \p{biact} of the BI action are supersymmetrized.

The rescaled $N=3$ superfield strengths  have the following
dimension:
\be
[W_{IK}]=[\bar{W}^{IK}]= -1~.\lb{D1}
\ee
The 4th
order superfield invariant lives in the same $N=3$ analytic superspace as the
free term \p{freehs}:
\bea
&&S_4={1\over32f^2}\int du d \zeta(^{33}_{11})
(W_{23})^2 (\bar{W}^{12})^2\nn\\
&&={1\over2f^2}\int d^4x V^2\bar{V}^2+\ldots~,\lb{D3}
\eea
where we omitted the fermionic and scalar field terms.

Given the function $E(V^2\bar V^2)$ defined by Eq. \p{explE}, let us
introduce the new function $\hat E(V^2\bar V^2)$ by
\be
E(V^2\bar{V}^2)={1\over2}V^2\bar{V}^2\hat{E}(V^2\bar{V}^2)~,
\ee
with $\hat{E}(a)=1-a/4+O(a^2)$. Then the whole sequence of
higher order terms in the $N=3$ generalization of the BI-action,
including the previous 4th order term, can be written as a closed
expression in the analytic superspace,
\be
S_E={1\over32f^2}\int du
d\zeta(^{33}_{11})(W_{23})^2(\bar{W}^{12})^2
\hat{E}(A)~.\lb{nlbiact}
\ee
Here $A$ is the following real
analytic superfield:
 \bea
&&A={1\over2^{11}}(D^1)^2 (\bar{D}_3)^2
[D^{2\alpha} W_{12}D^2_\alpha W_{12}\bar{D}_{2\da}\bar{W}^{23}
\bar{D}_2^\da\bar{W}^{23}]=V^2\bar{V}^2+\ldots~,\lb{D7}\\
&&W_{12}=D^3_1W_{23}=
-4i\,\theta^{(\alpha}_1\theta^{\beta)}_2V_\ab+\ldots~,\q
\bar{W}^{23}=-D^3_1\bar{W}^{12}~.
\eea

Thus, we have obtained an $N=3$ generalization of the Born-Infeld action
using the off-shell Grassmann-analytic potentials $\Vot$ and $\Vth$
\be
S^{N=3}_{BI} = S_2 + S_E~.
\ee
The substitution of generic function $\hat E(A)= 1 +O(A)$ into
\p{nlbiact} yields $N=3$ superextensions of the self-dual nonlinear
deformations of Maxwell theory discussed in the previous Section.

Note that one can modify the superfield $A$ in \p{D7} by terms containing
$(D^2)^2W_{12}$ or $(\bar{D}_2)^2\bar{W}^{23}$ which vanish on the free
mass shell (recall \p{short}). These extra terms do not influence the $SU(3)$
singlet sector of the  bosonic action, but can prove to be relevant for
implementing additional spontaneously broken symmetries (see \cite{BIK} for
the role of similar terms in the $N=2$ BI action).

\setcounter{equation}{0}
\section{Concluding remarks}
In this paper we have constructed a minimal $N=3$
superextension of Born-Infeld theory as the novel non-trivial
example of off-shell self-interacting gauge theory in $N=3$
HSS. Like the standard $N=3$ gauge theory action, the off-shell
$N=3$ BI action can be written as an integral over the
$(4{+}6|8)$-dimensional analytic subspace of the full $N=3$ HSS. It
yields the bosonic BI action in a new unusual form involving
tensor auxiliary fields which are present in the off-shell $N=3$
gauge multiplet. This $N=3$ supersymmetry-inspired form of the BI
action can be generalized to encompass a wide class of self-dual
deformations of the Maxwell action. All such nonlinear actions
admit $N=3$ supersymmetrization.

We conclude with a few remarks and conjectures.

Our nonlinear terms in the $N=3$ BI-action \p{nlbiact} generate
higher-order corrections to the free equations \p{C5} or \p{free2},
so in the nonlinear $N=3$ BI theory one cannot use the
standard abelian representations $V^I_K=iD^I_K v$ or \p{cbrep1}
even on shell. It is an interesting
problem to explore how the standard superfield constraints
\p{B5}, \p{B6} describing the free on-shell $N=3$ gauge theory
can be generalized to the BI deformation of the latter constructed here.

A closely related problem is as follows. In \cite{BIK}, an $N=4$
superfield form of the equations of motion of the $N=4$ super BI theory
with the second non-linearly realized $N=4$ supersymmetry was
given. It was derived in the framework of nonlinear realization
of the properly central-charge extended $N=8$ supersymmetry with
the unbroken $N=4$ subgroup. The basic Goldstone $N=4$ superfield is
associated with the central charge generator and it is a
generalization of the standard $N=4$ gauge superfield strength.
The $N=4$ BI equations are a covariantization, with respect to the
non-linearly realized $N=8$ supersymmetry, of the standard superfield
constraints of $N=4$ Maxwell theory and are expected to describe a
type II super D3-brane in $D=10$ in a static gauge. It
seems that this approach could be directly extended to the $N=3$ case. One
should start from a nonlinear realization of $N=6$ supersymmetry, such that
$N=3$  supersymmetry is unbroken and a complex central charge $Z^{ik}$ is
present in the anticommutator of the broken and unbroken $N=3$
supercharges. Then one introduces the Goldstone-Maxwell superfields
$W_{ik}(z), \bar W^{ik}(z)$ as the coset parameters associated
with $Z^{ik}, \bar Z_{ik}$ and replaces the derivatives in the
on-shell $N=3$ superfield constraints \p{B5}, \p{B6} by those
covariantized with respect to the nonlinear realization of $N=6$
supersymmetry. The resulting equations, like in the $N=8 \rightarrow N=4$
case, can be expected to contain a ``disguised'' form of the $D=4$
BI equations \cite{BIK} (along with the Nambu-Goto type equations for
6 physical scalars). Just like the constraints (\ref{B5}), (\ref{B6}) are
equivalent to the standard $N=3$ Grassmann analyticity conditions for the
harmonic projections \p{cbrep1} of the superfield strengths, the BI
deformation of these constraints can be equivalent to some nonlinear version
of the Grassmann analyticity conditions. One might expect that the latter,
like in the non-deformed case, can be transformed to a sort of nonlinear
harmonic equations for the analytic harmonic potentials, and that for these
equations a proper off-shell action exists. It should be a modification of the
minimal $N=3$ BI action constructed here, such that it reveals a second hidden
non-linearly realized $N=3$ supersymmetry. It is an intriguing open question
how to find the explicit relation between the $N=6 \rightarrow N=3$ coset
superfield variables and the off-shell $N=3$ Grassmann-analytic superfield
strengths.

One can approach the same problem from the opposite side and try to find
hidden spontaneously broken supersymmetries in the above off-shell
$N=3$ BI action, or in its proper modifications. Using the
general formula for the variation of the action $S_2(\Vot,
\Vth)$,
\be
\delta S_2(\Vot,\Vth)~\sim~\int d\zeta(^{33}_{11}) du
\left(\Vot[(\Doh- \Dth\Dot)\delta V^2_3+(\Dth)^2\delta
V^1_2]+\mbox{h.c.}\right)=0~,
\ee
it is straightforward to show
that $S_2(\Vot,\Vth)$ is off-shell invariant with respect to the
following Goldstone-type transformation of the harmonic gauge potentials:
\bea
&&\delta_G V^1_2=-(\delta_G
V^2_3)^\dagger=[(\theta_2)^2u_3^k-(\theta_2
\theta_3)u_2^k]\bar{c}_k +(\theta_2)^2(\epsilon\theta_3)
+2u_1^k(\eta_k\theta_2)(\bar\theta^1)^2\nn\\
&&+(\theta_2)^2
[(\bar\eta^k\bar\theta^1)u^2_k-(\bar\eta^k\bar\theta^2)u^1_k]~,\lb{goldtr}
\eea
where $c^k, \bar{c}_k, \eta_k^\alpha, \epsilon^\alpha,
\bar\eta^{k\da}$ and $\bar\epsilon^\da$ are additional 6 bosonic
and 16 Grassmann parameters. These transformations provide
shifts of the corresponding physical fields in the $N=3$ vector
multiplet and so can be treated as the lowest-order part of the
non-linearly realized symmetries. For the time being we do not
know whether a nonlinear generalization of the transformations
\p{goldtr} and the appropriate modification of the minimal $N=3$
BI action do exist. It is worth emphasizing that there are 16 shifting
fermionic symmetries. This can be viewed as an indication that
the hypothetical $N=3$ BI action with spontaneously broken
symmetries actually reveals the $N=8 \rightarrow N=4$ coset structure, with
one extra pair of broken and unbroken on-shell $N=1$
supersymmetries (and with all 16 physical fermions being
Goldstone ones). Thus it could provide a manifestly $N=3$
supersymmetric form of the action of the $N=8 \rightarrow N=4$ BI theory.
This conjecture is supported by the fact that the ordinary
$N=3$ gauge theory action coincides on shell with the action of $N=4$
gauge theory and so should show up one additional
on-shell supersymmetry. A technical problem tightly related to the issue
of hidden nonlinear symmetries and brane interpretation consists in
examining the 6 physical bosons sector of the $N=3$ BI action and comparing
it with the  Nambu-Goto action of 3-brane in $D=10$.

Finally, it is worthwhile to mention that our $N=3$ BI action admits
a straightforward nonabelian extension along the same lines as in the $N=1$
and $N=2$ cases \cite{ketnoa}.
\vspace{1.9cm}

\noindent{\Large\bf Acknowledgements}\\

\noindent The authors are grateful to Emery Sokatchev for
enlightening discussions at the early stage of this study and to Sergey
Krivonos for a useful remark. This work was partially supported by the
grants RFBR 99-02-18417,  RFBR-CNRS 98-02-22034, INTAS-00-0254, NATO Grant
PST.CLG  974874  and PICS Project No. 593.
\vspace{1cm}

\def\theequation{A.\arabic{equation}}
\setcounter{equation}0
\section*{\lb{F}Appendix A. $SU(3)/U(1)\times U(1)$ harmonics}

The $SU(3)/U(1)\times U(1)$ harmonics \cite{GIK2,GIO}  form an
$SU(3)$ matrix $u^I_i$ and are defined modulo the group
$U(1)\times U(1)$ which acts on the index $I$
\be
u^1_i=u^{(1,0)}_i~,\qquad u^2_i=u^{(-1,1)}_i~,\q
u^3_i=u^{(0,-1)}_i~.\lb{F1} \ee
Here $i$ is the index of the
fundamental representation of $SU(3)$. The complex conjugated
harmonics $u^i_I=\overline{u^I_i}$ have the opposite $U(1)\times U(1)$
charges:
\be u_1^i=u^{i(-1,0)}~,\qquad u_2^i=u^{i(1,-1)}~,\q
u_3^i=u^{i(0,1)}~.\lb{F2} \ee
The harmonics satisfy the following relations:
\bea
u^I_i
u^i_J=\delta^I_J~, \quad u^I_i u^k_I =\delta^k_i~,\quad
\varepsilon^{ikl}u^{(1,0)}_i u^{(-1,1)}_k
u^{(0,-1)}_l=1~.\lb{F5} \eea
The $SU(3)$-covariant harmonic derivatives act on the harmonics
according to the rule \be
\partial^I_J u^K_i=\delta^K_J u^I_i~,\qq
\partial^I_J u^i_K=-\delta^I_K u^i_J~.\lb{F9}
\ee
The special $SU(3)$ conjugation $\sim $ of the harmonics is defined by
\be
\widetilde{u^1_i}=u^i_3~,\q\widetilde{u^3_i}=u^i_1~,\q
\widetilde{u^2_i}=-u^i_2~.\lb{F13b}
\ee
On the  harmonic projections of spinor coordinates
\be
\theta^\alpha_I=u_I^i\theta^\alpha_i~,\q\bar\theta^{\da I}=u^I_i
\bar\theta^{\da i} \ee
the $\sim$ conjugation acts in the
following way:
\be
\toa\leftrightarrow\btha~,\qq
\tta\leftrightarrow-\btta~, \qq \theta^\alpha_3 \leftrightarrow \bar
\theta^{1\dot\alpha}~. \ee
The conjugation rules of harmonic derivatives are as follows:
\be
\widetilde{\Doh f}=-\Doh\tilde{f}~,\q\widetilde{\Dot
f}=\Dth\tilde{f}~. \lb{F13c}
\ee

The analytic superspace $H(4{+}6|8)$ is parametrized by the
coordinates $\{\zeta, u\}$, where
\bea
&& \zeta^M \equiv \{ x^\adb_\A=x^\adb
+4i(\toa \btob -\tha \bthb),~ \tta,~\tha,~\btoa,~\btta \}~,
\lb{F14b}\\
&&\delta x^\adb_\A=4i(\tta u^2_k+2i\tha u^3_k)\beka-4i\eka(2\btoa
u_1^k+ \btta u_2^k)~,\lb{F14}
 \eea
and it is closed (i.e., real) under the generalized
conjugation. In these coordinates the spinor and harmonic derivatives
have the following explicit form:
\bea
&&D^1_\alpha =\poa~,\qq \bar{D}_{3\da} =-\bpha~,  \nn \\
&&\bDoa=-\bpoa -4i\tob\pbda~,\q D^3_\alpha =\pha
+4i\bthb\padb~,\nn \\ &&D^2_\alpha =\pta
+2i\bttb\padb~,\q \bDta=-\bpta
-2i\ttb\pbda~,\lb{F18} \\
&&D^1_2=\pot
+2i\tta\btob\partial_\adb -\tta\poa + \btoa\bpta~,
\nn \\ &&D^2_3=\pth +2i\tha\btta\partial_\adb -\tha\pta
+ \btta\bpha~, \nn \\ &&D^1_3 =\poh +4i\tha\btoa\partial_\adb
-\tha\poa + \btoa\bpha~, \nn\\
&&D_1^2=\pto
-2i\toa\bttb\partial_\adb -\toa\pta + \btta\bpoa~,
\nn \\ &&D_2^3=\pht -2i\tta\btha\partial_\adb
-\tta\pha + \btha\bpta~, \nn \\ &&D_1^3 =\pho
-4i\toa\btha\partial_\adb -\toa\pha + \btha\bpoa~. \lb{F23b}
\eea
where $\padb= \partial/\partial x_\A^{\alpha\dot\beta}$.

The Grassmann and harmonic measures of integration over the $N=3$
analytic harmonic superspace are normalized so that
\be
\int
d^8\theta(^{33}_{11})(\theta_2)^2(\theta_3)^2(\bar\theta^1)^2(\bar\theta^2)^2
=1~,\q\int du=1~.\lb{gme2}
\ee

\renewcommand\theequation{B.\arabic{equation}} \setcounter{equation}0
\section*{\lb{Tu}Appendix B. Relation between spinor and vector\\
representations}
\bea
&& x^{\alpha\dot\beta} = (\tilde\sigma_m)^{\beta\dot\alpha} x^m~, \quad
\partial_{\alpha\dot\beta} = \partial/\partial x^{\alpha\dot\beta} = {1\over 2}
\,(\sigma^m)_{\alpha\dot\beta}\partial/\partial x^m~, \quad A_{\alpha\dot\beta}
= (\sigma^m)_{\alpha\dot\beta} A_m~,  \nn \\
&&F_\ab={1\over2}\left( \partial^\db_\alpha
A_\bdb+ \partial^\db_\beta A_\adb\right)~,\q
\bar{F}_{\da\db}={1\over2}\left(\partial^\beta_\da A_\bdb
+ \partial^\beta_\db A_\bda\right)~, \nn\\
&&{\cal F}_{mn}=\partial_mA_n-
\partial_n A_m = {i\over 2}\,F^{\alpha\beta}(\sigma_{mn})_{\alpha\beta} -
{i\over 2}\, \bar F^{\dot\alpha\dot\beta} (\tilde
\sigma_{mn})_{\dot\alpha\dot\beta}~, \nn \\
 &&({\cal F}_{mn})^2=2\left( F^2+ \bar{F}^2\right)~,\q
{1\over2}\,\varepsilon^{mnpq}{\cal F}_{mn} {\cal F}_{pq}=-2i \left( F^2
-\bar{F}^2 \right)~. \lb{vecsp}
\eea

We use $\eta_{mn} = \mbox{diag}(1,-1,-1,-1)$ and the standard conventions of
the two-component spinor formalism \bea
&& (\sigma_m)_{\alpha\dot\beta} = (1, \vec\sigma)_{\alpha\dot\beta}~, \q
(\tilde\sigma_m)^{\dot\beta\alpha} =
\varepsilon^{\alpha\beta}\varepsilon^{\dot\beta\dot\alpha}
(\sigma_m)_{\beta\dot\alpha}~,
\nn \\
&& (\sigma_{mn})_{\alpha\beta} ={i\over 2}\left(\sigma_m\tilde\sigma_n -
\sigma_n\tilde\sigma_m\right)_{\alpha\beta}~, \q
(\tilde\sigma_{mn})_{\dot\alpha\dot\beta} ={i\over
2}\left(\tilde\sigma_m\sigma_n -
\tilde\sigma_n\sigma_m\right)_{\dot\alpha\dot\beta}~, \nn \\
&& \varepsilon_{12} = \varepsilon_{\dot{1}\dot{2}} = -\varepsilon^{12} =
-\varepsilon^{\dot{1}\dot{2}} = 1~, \q
\varepsilon^{\alpha\beta}\varepsilon_{\beta\gamma} = \delta^\alpha_\gamma~, \;
\varepsilon^{\dot\alpha\dot\beta}\varepsilon_{\dot\beta\dot\gamma} =
\delta^{\dot\alpha}_{\dot\gamma}~.
\eea

\end{document}